# Massive parallel generation of indistinguishable single photons via polaritonic superfluid to Mott-insulator quantum phase transition

Neil Na[1] and Yoshihisa Yamamoto[2,3]

[1]*Intel Corporation, 2200 Mission College Blvd., Santa Clara, CA, 95054, USA*

[2]*E. L. Ginzton Laboratory, Stanford University, Stanford, California 94305, USA*

[3]*National Institute of Informatics, Hitotsubashi, Chiyoda-ku, Tokyo 101-8430, Japan*

We propose the superfluid to Mott-insulator quantum phase transition in an array of exciton-polariton traps can be utilized for massive parallel generation of indistinguishable single photons. By means of analytical and numerical methods, the device operations and system properties are studied using realistic experimental parameters. Such a deterministic, fault-tolerant, massive parallel generation may open up a new perspective in photonic quantum information processing.

*Introduction* – Generation of indistinguishable single photons from a large number of independent emitters is essential in many recent proposals of scalable quantum information processing [1]. Preliminary experimental efforts have demonstrated the feasibility in systems such as trapped atoms and ions [2], as well as impurity-bound excitons [3]. However, in these works where single photons are generated by the radiative decay of spatially independent emitters pumped by incoherent optical excitation, only up to two independent single-photon sources can be prepared. A collective generation of many indistinguishable single photons simultaneously still remains out of reach.

In this paper, we propose a novel approach to generate indistinguishable single photons in a massive parallel fashion. More importantly, the system can be deterministically controlled and the impact of inevitable fabrication disorder is shown to have limited influence. The basic idea is to load a dilute gas of exciton-polaritons [4] in a periodic potential traps, and drive the system across the superfluid (SF) to Mott-insulator (MI) quantum phase transition (QPT) by modulating the photon-exciton frequency detuning. Indistinguishable single photons can then be triggered independently in the MI phase by the radiative decay of exciton-polaritons. As a consequence, massive amount of indistinguishable single photons can be obtained parallelly in this scheme. Such a polaritonic QPT from a SF to MI state was predicted recently in a variety of solid-state systems, such as a cavity array containing four-level atomic ensembles in an EIT configuration, single-atom cavity QED array, and excitonic cavity QED array [5,6]. The existence of Bose-glass phase due to system disorder was also studied [6]. In the following paragraphs, we'll discuss in details the generation scheme including device operations and system properties. This deterministic, fault-tolerant, massive parallel generation of indistinguishable single photons is essential for applications in scalable quantum computation and communication, and could potentially find new applications in photonic quantum information processing.

*Experimental Setup* – Fig. 1 shows a schematic plot of the proposed device. A single GaAs quantum well (QW) is embedded in a half-wavelength $Al_xGa_{1-x}As$ optical cavity layer, which is sandwiched in between the upper and lower distributed-bragg-reflectors

(DBR). The optical cavity layer thickness is spatially modulated by etching small mesas that serve as photon trapping centers. Details of the formation of these three-dimensionally confined microcavities can be found in Ref. [7], and here we take advantage of the results that they can be treated as single-mode cavities in the following. Metal gates are fabricated on top to apply a vertical electric field so that the photon-exciton frequency detuning can be controlled by quantum-confined Stark effect (QCSE) [8]. Photons and excitons in this system are strongly coupled to each other, and their normal modes are defined as polaritons. The dynamics of such an array of exciton-polariton traps can be described by the Bose-Hubbard model (BHM) with a system-reservoir coupling, which will be discussed in depth in the next paragraph. The lower DBR is made thicker than the upper DBR to enforce single-side cavity emission. The modulated planar microcavities inherit circular symmetry and are suitable for coupling to down-stream fiber-optics with high collection efficiency. Note that although a specific setup is discussed in this paper to validate our experimental proposal, the concept can apply to different variation of materials, type of cavities, and control of detuning.

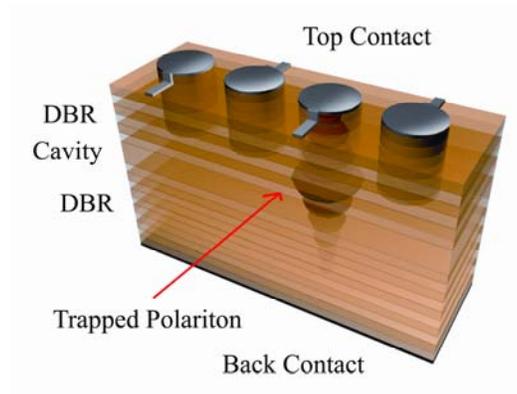

**Figure 1.** A schematic plot of the proposed device for massive parallel generation of indistinguishable single photons.

*System Hamiltonian* – The system Hamiltonian is given by

$$H = \sum_{c=a,b} \int d\mathbf{r} \Psi_c^\dagger(\mathbf{r}) \left( \frac{-\nabla^2}{2m_c} + V_c(\mathbf{r}) \right) \Psi_c(\mathbf{r}) + g' \int d\mathbf{r} \Psi_a^\dagger(\mathbf{r}) \Psi_b(\mathbf{r}) + H.c.$$
$$+ \frac{u'}{2} \int d\mathbf{r} \Psi_b^\dagger(\mathbf{r}) \Psi_b^\dagger(\mathbf{r}) \Psi_b(\mathbf{r}) \Psi_b(\mathbf{r}) - \Delta g' \int d\mathbf{r} \Psi_b^\dagger(\mathbf{r}) \Psi_a^\dagger(\mathbf{r}) \Psi_b(\mathbf{r}) \Psi_b(\mathbf{r}) + H.c. \quad (1)$$
$$+ \int d\mathbf{r} f'(\mathbf{r},t) e^{-i\nu t} \Psi_a^\dagger(\mathbf{r}) + H.c.$$

where the field operators $\Psi_a$ and $\Psi_b$ refer to cavity photon and QW exciton. The first term in (1) represents the free Hamiltonians of trapped photons and excitons. The second through fifth terms correspond to photon-exciton coupling, exciton-exciton repulsion, reduction of excitonic dipole moment, and external laser coupled to cavity mode, respectively. Since the effective mass of a QW exciton is much larger than that of a cavity photon, it is appropriate to define a single-mode exciton operator $b_i$ that features the same wavefunction as of single-mode photon operator $a_i$ [9]. By doing so, (1) can be rewritten as

$$H = \omega_a \sum_i a_i^\dagger a_i + t \sum_{<ij>} a_i^\dagger a_j + \omega_b \sum_i b_i^\dagger b_i + g \sum_i \left( a_i^\dagger b_i + H.c \right)$$
$$+ \frac{u}{2} \sum_i b_i^\dagger b_i^\dagger b_i b_i - \Delta g \sum_i \left( b_i^\dagger a_i^\dagger b_i b_i + H.c \right) \quad (2)$$
$$+ f(t) \sum_i \left( a_i^\dagger e^{-i\nu t} + H.c. \right)$$

$\omega_a$ and $\omega_b$ are the site cavity photon and QW exciton energies. $t$ is the photon tunneling energy determined by the overlapping of nearest-neighbor cavity fields. $g$ is the photon-exciton coupling constant. $u$ and $\Delta g$ are energies that correspond to the exciton-exciton repulsion and the reduction of excitonic dipole moment. $f(t)$ and $\nu$ are the external laser amplitude and energy. Next, we define the upper polariton (UP) and lower polariton (LP) operators $q_i$ and $p_i$ as a linear superposition of $a_i$ and $b_i$ with appropriate Hopfield coefficients A and B [10]. The system master equation for LPs in the rotating frame of the external laser is derived as

$$\frac{d\rho}{dt} = \frac{1}{i}[\tilde{H}, \rho] - \frac{\Gamma}{2}\sum_i (\rho p_i^\dagger p_i + p_i^\dagger p_i \rho - 2 p_i \rho p_i^\dagger) \qquad (3)$$

under rotating wave approximation, where

$$\tilde{H} = -\Delta \sum_i p_i^\dagger p_i - J\sum_{<ij>} p_i^\dagger p_j + \frac{U}{2}\sum_i p_i^\dagger p_i^\dagger p_i p_i + F(t)\sum_i (p_i^\dagger + p_i). \qquad (4)$$

UP dynamics are discarded because the external laser selectively pumps the LPs. $\Delta$ is the energy difference between the external laser and the trapped LPs. $J$ is the LP tunneling energy and is equal to $tA^2$. $U$ is the LP-LP interaction energy and is equal to $uB^4 + 4\Delta g B^3 A$. Assuming an infinite potential barrier with area $S$ for photon trapping, $u$ can be calculated by $\sim 2.2 E_B \cdot \pi a_B^2 / S$ due to fermionic exchange interaction, and $\Delta g$ can be calculated by $\sim 4g \cdot \pi a_B^2 / S$ due to phase space filling and fermionic exchange interaction [11]. $E_B$ and $a_B$ are the 1s exciton binding energy and Bohr radius. Let $S = \pi(\lambda/2)^2$ where $\lambda = 222$ nm (the emission wavelength of a 10 nm GaAs QW divided by GaAs refractive index at 4 K), $u$ and $\Delta g$ are derived as 200 and 90 μeV, given $E_B = 10$ meV, $a_B = 10$ nm, and $g = 2.5$ meV. $F(t)$ is equal to $f(t)A$. $\Gamma$ is the LP decay rate and is equal to $A^2 Q/\omega_a + B^2/\tau_b$. Cavity $Q$ factor equal to $10^6$ and QW exciton lifetime $\tau_b$ equal to 0.5 ns are used. Note that because the acoustic phonon-polariton scattering time exceeds 1 ns for zero in-plane momentum regime at 4 K, and the polariton-polariton scattering is negligible for LP density smaller than $10^{10}$ cm$^{-2}$, our system decoherence is expected to be limited by the radiative process.

For an ideal 1D system with unit filling, the critical point of BHM calculated by quantum Monte-Carlo simulation is $U/J_c \sim 2.04$ [12]. If we assume the polariton lifetime is long enough compared to all other time scales, this condition of QPT can be reasonably applied in our system [6]. In Fig. 2, we plot $U/J$ as a function of photon-exciton frequency detuning $\delta=\omega_a-\omega_b$, given different $t$ values that are determined by the inter-cavity distance. It is found that the critical point can be reach by modulating a negative $\delta$ (red detuning) into a positive $\delta$ (blue detuning), i.e., changing from a photon-like polariton into an exciton-like polariton. This is physically expected, because, an exciton-like polariton features larger $U$ (due to exciton nonlinearity) and at the same time smaller $J$ (due to photon tunneling).

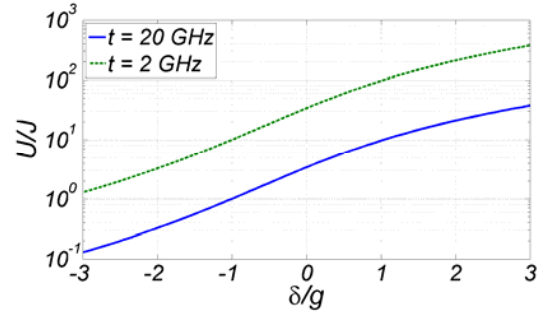

**Figure 2.** The ratio of polariton interaction energy to polariton tunneling energy plotted as a function of photon-exciton frequency detuning. Two photon tunneling energies, 2 GHz and 20 GHz, are examined; for $t=20$ GHz, it corresponds to ~2 μm of inter-cavity distance assuming a Gaussian-like cavity field.

*Device operations and system properties* – Numerical simulations are performed by discretizing (3) in the time domain where the matrix representations of all operators are constructed. Due to the huge increase of Hilbert space size with cavity number, we choose six one-dimensional coupled cavities with periodic boundary conditions. The sharp SF to MI QPT is smeared in such a finite number of cavities [5], but suffices to prove the operation principle of our proposal.

The device operation procedures are shown in Fig. 3 (a) and (b) for the odd and even numbered cavities, respectively. The system is initially (at 0 ps) prepared in a photon-like SF state where $U/J \sim 0.13$, which is realized by a large red photon-exciton frequency detuning $\delta = -3g$ and an numerical excitation condition

$$\frac{1}{N!}(\frac{1}{\sqrt{N}}\sum_{i=1}^{N} p_i^\dagger)^n \rho_o (\frac{1}{\sqrt{N}}\sum_{i=1}^{N} p_i)^n . \qquad (5)$$

$N$ is the cavity number, $n$ is the polariton number, and $\rho_o$ is the density matrix of vacuum, respectively. Note that (5) excites on-average one polariton per cavity that hops randomly in the coupled cavities. Experimentally, this can be achieved by controlling the external laser coupled to the cavity mode with appropriate pulse amplitude and width, because a coherent field excitation can mimic the initialization condition (5) in the thermodynamic limit ($N \gg 1$). Then, by using QCSE (from 0 to 200 ps), the QW exciton energy is lowered by the applied vertical electric field so that $\delta$ is switched from $-3g$ to $4g$, i.e., into an exciton-like MI state where $U/J \sim 35$. A single LP is localized in each cavity due to the dominance of polariton-polariton interaction over nearest-neighbor tunneling. The shape of electrical switch pulse follows a hyperbolic tangential function with switching speed equal to 10 GHz, which is chosen to perform an adiabatic transition during this time window. Finally, while $\delta$ of the even numbered cavities stay at $4g$, $\delta$ of the odd numbered cavities are switched rapidly at the speed of 1 THz back to $-4g$ (at 200 ps). Single photon emissions are now triggered from the odd number cavities, and the purpose of such a selective switching will be explained shortly. Note that $\tau_b$ and $g$ are independent of $\delta$ because the lifetime and oscillator strength of a QW exciton barely change for the range of vertical electric field used in the above $\delta$ switching [8].

The use of selective switching at 250 ps is under three considerations. First, the quantum efficiency of generating single photons [13]

$$\eta = \int <p_i^\dagger(t) p_i(t)> A^2(t) \frac{Q}{\omega_a} dt \tag{6}$$

should be maximized so that a polariton decay is mostly directed to the cavity mode. Switching the exciton-like LPs back to the photon-like LPs achieves this goal. Second, if all of the cavities are switched back to a large red detuning regime, rapid tunneling process with $J/\Gamma \sim 194$ readily destroys the deterministic single polariton decay from individual site (simulation results not shown). Instead, in the present selective switching, only the LPs in the odd numbered cavities are switched back to a large red detuning regime so that the neighboring site energy mismatch effectively cuts off the unwanted tunneling events. Finally, the frequency of the emitted single photons is tuned away from that of the external laser. Using a narrow band-pass frequency filter, clean output signal can be selected out.

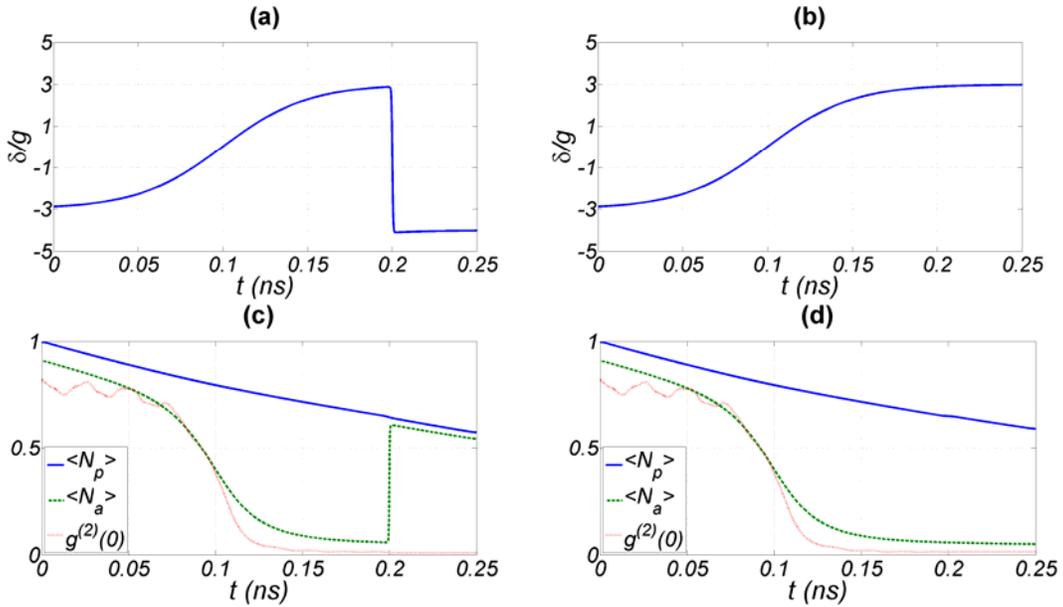

**Figure 3.** The device operation procedures and dynamics in the odd (a) (c) and even (b) (d) numbered cavities plotted as a function of time. In (a) (b), the solid-blue line corresponds to the electrical switch pulse. In (c) (d), the solid-blue, dashed-green, and dotted-red lines correspond to the average LP number, average photon number, and LP second-order coherence, respectively.

The dynamics of the odd and even numbered cavities are shown in Fig. 3 (c) and (d), respectively. During the adiabatic $\delta$ switching, the normalized zero-delay second-order coherence function $g^{(2)}(0)$ starts with ~0.81 at 0 ps due to injecting 6 photon-like LPs that hop randomly in the coupled cavities, and subsequently drops to ~0.01 at 200 ps due to localizing 1 exciton-like LP in each cavity. This strongly antibunching behavior indicates the crossing of SF to MI boundary. The effect of selective switching can be seen from the sharp increase of the average photon number $<N_a>$ in the odd numbered cavities. $\eta$ of the single photon emissions is ~79.5% in Fig. 3, and can be further maximized by carefully designing the switch pulse shape. The ultimate physical limit of $\eta$ comes from how large $U$ or $J_c$ can be and therefore how fast an adiabatic $\delta$ switching may use.

To further understand the system dynamics, we define two parameters: the far-field optical interference visibility [14]

$$V(t) \equiv \frac{<n_a(t)>_{max} - <n_a(t)>_{min}}{<n_a(t)>_{max} + <n_a(t)>_{min}} \qquad (7)$$

where

$$n_a(t) = \frac{1}{N} \sum_{m,n \in Z} a_m^\dagger(t) a_n(t) e^{i(m-n)\varphi}, \qquad (8)$$

and the single photon indistinguishability [15]

$$I(t) \equiv 1 - \frac{<c_1^\dagger(t)c_3^\dagger(t)c_3(t)c_1(t)>}{<c_1^\dagger(t)c_1(t)><c_3^\dagger(t)c_3(t)>} \qquad (9)$$

where

$$\begin{pmatrix} c_1 \\ c_3 \end{pmatrix} = \frac{1}{\sqrt{2}} \begin{pmatrix} 1 & 1 \\ -1 & 1 \end{pmatrix} \begin{pmatrix} a_1 \\ a_3 \end{pmatrix}. \qquad (10)$$

$\varphi$ is the optical phase difference between each cavity. The visibility $V(t)$ measures the first-order phase coherence through the far-field optical interference contrast. The indistinguishability $I(t)$ measures the identicality of the two photons emitted simultaneously from cavities number 1 and 3 through the Hong-Ou-Mandel interferometer. These quantities are plotted as a function of time in Fig. 4, where all the parameter settings are the same as in Fig. 3. As expected, $I(t)$ rises to nearly 1 at 200 ps, suggesting the generation of indistinguishable single photons. On the other hand, $V(t)$ drops from 1 to 0.29 rather than 0 at the end. Note that the finite visibility implies a residual tunneling effect, which is a direct reflection of the non-unity $\eta$ caused by the polariton loss through radiative decay [5] before triggering the emissions of indistinguishable single photons. This is confirmed by artificially increasing $Q$ and $\tau_b$ by an order of magnitude, and we find $V(t)$ further drops to 0 while $I(t)$ still rises to nearly 1.

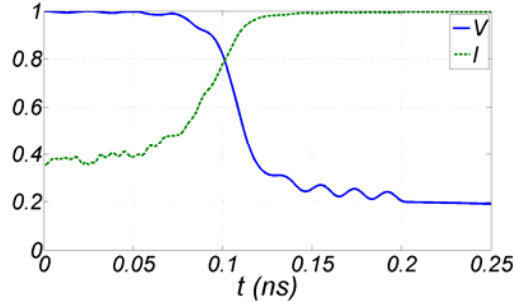

**Figure 4**. The far-field optical interference visibility (solid-blue line) and the single photon indistinguishability (dashed-green line) plotted as a function of time. Only odd

numbered cavities are taken into calculations. The slight oscillations in both parameters indicate the weak non-adiabaticity due to the used of a moderately fast $\delta$ switching.

*Experimental considerations* – Unlike ultracold atoms in an optical lattice where an extremely clean experimental environment can be prepared, disorder due to the fabrication error of solid-state devices is unavoidable. One important benefit of our proposal is its robustness against such an imperfection: first, the site energy disorder can be manually addressed and compensated by the QCSE, which alleviates the inhomogeneity seen by the LPs. Second, since the system is prepared initially in a SF state, the site energy disorder is effectively reduced by roughly a factor of $d/J$. Compared to the other deterministic generation scheme such as photon blockade (PB) effect where the bandwidth of a pumped $\pi$ pulse cannot spectrally well overlap the inhomogeneous LP site energies, the initialization of LP population in our scheme can be much more uniform. Note that the increase of a pumped $\pi$ pulse bandwidth in the PB scheme to improve the spectral coupling is not allowed because a second LP is then excited and breaks down the PB principle. Based on these two benefits, our proposal can largely overcome the site energy disorder such as inhomogeneous broadening of cavity photons and QW excitons, and therefore promises a practical path toward massive parallel generation of indistinguishable single photons.

The required temperature for our proposal determines the feasibility of a laboratory demonstration. To avoid particle-hole excitation in a MI state, we need a thermal energy $KT$ to be much smaller than $U$. Suppose $KT$ is an order of magnitude smaller than $U$, $T\sim0.2$ K is in general needed. This increases the experimental difficulties because a dilute refrigerator must be used. Nevertheless, the proposed system operation is based on a coherent spectroscopic technique and a serious thermalization effect kicks in only when the LPs are exciton-like, which lasts shorter than 100 ps during the device operation procedures (see Fig. 3 (c) (d)). Such a number is smaller than the typical thermalization time in an exciton-polariton system at 4 K, and in this sense we may really probe the zero-temperature quantum dynamics shown above.

*Generation of polarization-entangled photon pairs* – So far we have neglected the spin of a LP by assuming a circularly-polarized external laser is used for optical pumping. It is possible to generate polarization-entangled photon pairs via the QPT from a SF to MI state if the two spin species are simultaneously injected. Our scheme is illustrated in Fig. 5. Initially, a linearly-polarized external laser injects on-average two LPs per cavity at a large $J/U$, which forms a photon-like SF. Subsequent adiabatic $\delta$ switching sweeps the system into an exciton-like MI with two localized LPs in each cavity. While more studies are required to establish the exact phase diagram of spin-dependent interacting polaritons, we argue the ground state of the proposed scenario is a collection of two opposite-spin LPs occupying the same site. This is due to fact that the electron (and hole) component of an exciton must satisfy Paul-exclusive principle, so that the followed emissions are similar to the biexciton emissions in a semiconductor quantum dot [17,18]. By using the selective switching as described above, two-photon cascaded emission is triggered where the anticorrelation of LP spins is translated to the circularly-polarized states of photons. A maximally polarization-entangled photon pair $(|\sigma^+\rangle_1|\sigma^-\rangle_2+|\sigma^-\rangle_1|\sigma^+\rangle_2)/\sqrt{2}$ can be obtained, where subscripts 1 and 2 refer to the first and second photon emitted that have an energy difference equal to $U$.

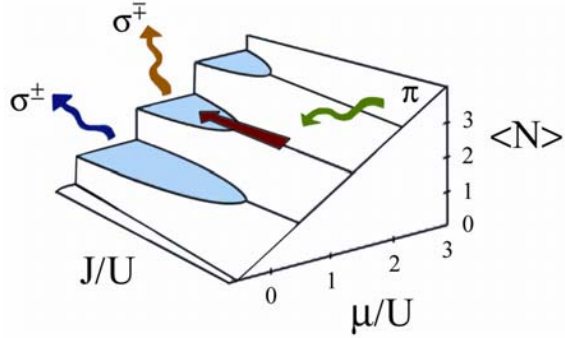

**Figure 5**. Phase diagram of the BHM. The system is first pumped by a linearly-polarized ($\pi$) external laser at a large $J/U$ to $<N>=2$, and then followed by an adiabatic $\delta$ switching indicated by the red arrow to cross the SF-MI boundary. Subsequent selective switching triggers polarization-entangled photon pairs that are circularly-polarized ($\sigma$).

*Summary* – we have shown how to harness the polaritonic QPT from a SF to MI state to deterministically generate indistinguishable single photons. The system robustness against site energy disorder paves a practical route to nonclassical photon generation in a massive parallel fashion. A variety of other applications as photon number eigenstate interferometer [19] and subwavelength quantum lithography [20], can benefit from our proposed scheme.

*Acknowledgment* – One of the authors (N. N.) was partially supported by Mediatek Fellowship. We would like to thank Q. Zhang for useful discussion. This work was supported by SORST program of Japan Science of Technology Corporation (JST), and the University of Tokyo (CINQIE).